\documentclass[pdflatex,sn-nature]{sn-jnl}

\usepackage{graphicx}%
\usepackage{multirow}%
\usepackage{amsmath,amssymb,amsfonts}%
\usepackage{amsthm}%
\usepackage{mathrsfs}%
\usepackage[title]{appendix}%
\usepackage{xcolor}%
\usepackage{textcomp}%
\usepackage{manyfoot}%
\usepackage{booktabs}%
\usepackage{algorithm}%
\usepackage{algorithmicx}%
\usepackage{algpseudocode}%
\usepackage{listings}%

\theoremstyle{thmstyleone}%
\theoremstyle{thmstyletwo}%

\theoremstyle{thmstylethree}%

\begin{document}

\title[Article Title]{Deterministic Frequency--Domain Inference of Network Topology and Hidden Components via Structure--Behavior Scaling}


\author[1]{\fnm{Xiaoxiao} \sur{Liang}}\email{liang\_x@mail.ustc.edu.cn}

\author*[1]{\fnm{Tianlong} \sur{Fan}}\email{tianlong.fan@ustc.edu.cn}

\author*[1]{\fnm{Linyuan} \sur{Lü}}\email{linyuan.lv@ustc.edu.cn}

\affil[1]{%
  \orgdiv{School of Cyber Science and Technology},%
  \orgname{University of Science And Technology of China},%
  \orgaddress{%
    \city{Hefei},%
    \state{Anhui},%
    \country{China}%
  }%
}


\abstract{Hidden interactions and components in complex systems—ranging from covert actors in terrorist networks to unobserved brain regions and molecular regulators—often manifest only through indirect behavioral signals. Inferring the underlying network structure from such partial observations remains a fundamental challenge, particularly under nonlinear dynamics. We uncover a robust linear relationship between the spectral strength of a node’s behavioral time series under evolutionary game dynamics and its structural degree, \( S \propto k \), a structural–behavioral scaling that holds across network types and scales, revealing a universal correspondence between local connectivity and dynamic energy. Leveraging this insight, we develop a deterministic, frequency-domain inference framework based on the discrete Fourier transform (DFT) that reconstructs network topology directly from payoff sequences—without prior knowledge of the network or internal node strategies—by selectively perturbing node dynamics. The framework simultaneously localizes individual hidden nodes or identifies all edges connected to multiple hidden nodes, and estimates tight bounds on the number of hidden nodes. Extensive experiments on synthetic and real-world networks demonstrate that our method consistently outperforms state-of-the-art baselines in both topology reconstruction and hidden component detection. Moreover, it scales efficiently to large networks, offering robustness to stochastic fluctuations and overcoming the size limitations of existing techniques. Our work establishes a principled connection between local dynamic observables and global structural inference, enabling accurate topology recovery in complex systems with hidden elements.}




\maketitle

\section{Introduction}\label{sec1}

Complex networks provide a powerful framework for analyzing the structural and dynamical properties of systems across diverse domains, including biological networks \citep{bioinfer2007,geneinfer2009}, neural systems \citep{brain2008}, social communities \citep{gao2009complex,memon_larsen_hicks_harkiolakis_2008}, and critical infrastructures \citep{introtoflowsys,li2017reconstruction}. Despite significant advances in our understanding of how network topology influences system dynamics, accurately inferring the underlying network structure from observed behaviors remains a fundamental challenge, particularly when interactions are nonlinear and unobservable nodes—known as hidden nodes—exist \citep{han_shen_wang_di_2015,ha2020deep}. Hidden nodes and their interactions are ubiquitous and pose significant challenges in practice; for instance, identifying covert leaders in terrorist networks is crucial for intelligence analysis \citep{memon_larsen_hicks_harkiolakis_2008}, detecting latent regulatory factors is vital in biological systems \citep{bioinfer2007,geneinfer2009}, and uncovering unobserved brain regions is essential for accurate neurological diagnoses \citep{brain2008}.

Numerous methods have been developed for network inference, including causality analysis \citep{sugihara_may_ye_hsieh_deyle_fogarty_munch_2012}, matrix inversion techniques \citep{shandilya_timme_2011}, compressive sensing approaches \citep{wang_yang_lai_kovanis_grebogi_2011}, and machine learning frameworks \citep{zhang_zhao_liu_wang_tao_xin_zhang_2019}. However, these approaches face fundamental limitations when applied to discrete-time, nonlinear interactions commonly observed in social, biological, and economic systems \citep{evolutionary1998,evolutionary2006}. Most notably, they assume prior knowledge of the explicit form of the node dynamics—such as continuous oscillator-type systems governed by coupled differential equations—or rely heavily on linear statistical assumptions. They also require full observability and prior knowledge of the network size, making them unsuitable for scenarios involving hidden or unobserved nodes. Furthermore, their applicability is typically restricted to small-scale networks due to prohibitive computational complexity, and their performance often degrades in the presence of noise and stochastic fluctuations inherent in real-world data.

To overcome these challenges, this study introduces a deterministic, frequency-domain framework leveraging the DFT to infer network topology directly from behavioral time series generated by evolutionary game interactions, without requiring knowledge of the underlying node dynamics or full observability. Central to our method is the discovery of a robust linear relationship between a node’s spectral strength of its payoff dynamics and the structural degree, described succinctly by $S \propto k$. This linear scaling relationship holds consistently across diverse network types and scales, constituting a fundamental theoretical contribution that bridges local dynamic observables with global structural inference. By selectively perturbing individual nodes and analyzing frequency-domain changes in their neighbors' payoff sequences, our framework not only reconstructs the network topology with high accuracy but also effectively identifies hidden nodes and their external connections, even in the absence of prior topological knowledge or internal node-level strategies.

Extensive experiments conducted on both synthetic models and real-world empirical networks demonstrate the robustness, scalability, and superior performance of our proposed method compared to state-of-the-art baseline techniques. The method achieves consistently high reconstruction accuracy, efficiently scales to large systems, and maintains effectiveness under varying stochastic perturbations. Critically, the capability to simultaneously localize single hidden nodes, identify external edges connected to multiple hidden nodes, and estimate tight bounds on hidden node counts addresses the practical challenges posed by partially observable complex networks.

Collectively, this work not only advances our theoretical understanding of how structural information manifests within nonlinear behavioral dynamics but also provides a powerful, universally applicable inference framework suitable for addressing critical problems in network science, biology, neuroscience, and security applications, where hidden interactions profoundly impact system dynamics and outcomes.

\section{Results}\label{results}

We systematically evaluate the effectiveness of our DFT-based inference framework through a series of experiments designed to test its theoretical foundation, reconstruction accuracy, and capacity to reveal hidden network elements. The results are organized into three parts. First, we validate the theoretical linear relationship between node strength and degree using payoff dynamics derived from evolutionary games. Next, we assess the method's performance in reconstructing the underlying network topology across both synthetic and real-world networks, comparing it against several representative baselines. Finally, we demonstrate the ability of the proposed approach to detect hidden nodes and estimate their external connectivity.

All experiments are conducted under consistent modeling conditions, without requiring prior knowledge of the network structure or node-level dynamics. This generality allows the framework to be applied across diverse network types, ranging from well-controlled synthetic graphs to empirical systems with strong structural heterogeneity and incomplete observations.

\subsection{Validation of Structure–Behavior Linearity}
\label{subsec:linearity}

A core theoretical principle of our framework is that the frequency-domain strength of a node's behavioral time series scales linearly with its structural degree, i.e., $S \propto k$. To empirically validate this scaling law, we simulate evolutionary game dynamics over three network models—ER, BA, and WS—across a range of network sizes and game parameters. 

\begin{figure*}[htbp]
	\centering
	\includegraphics[scale=0.5]{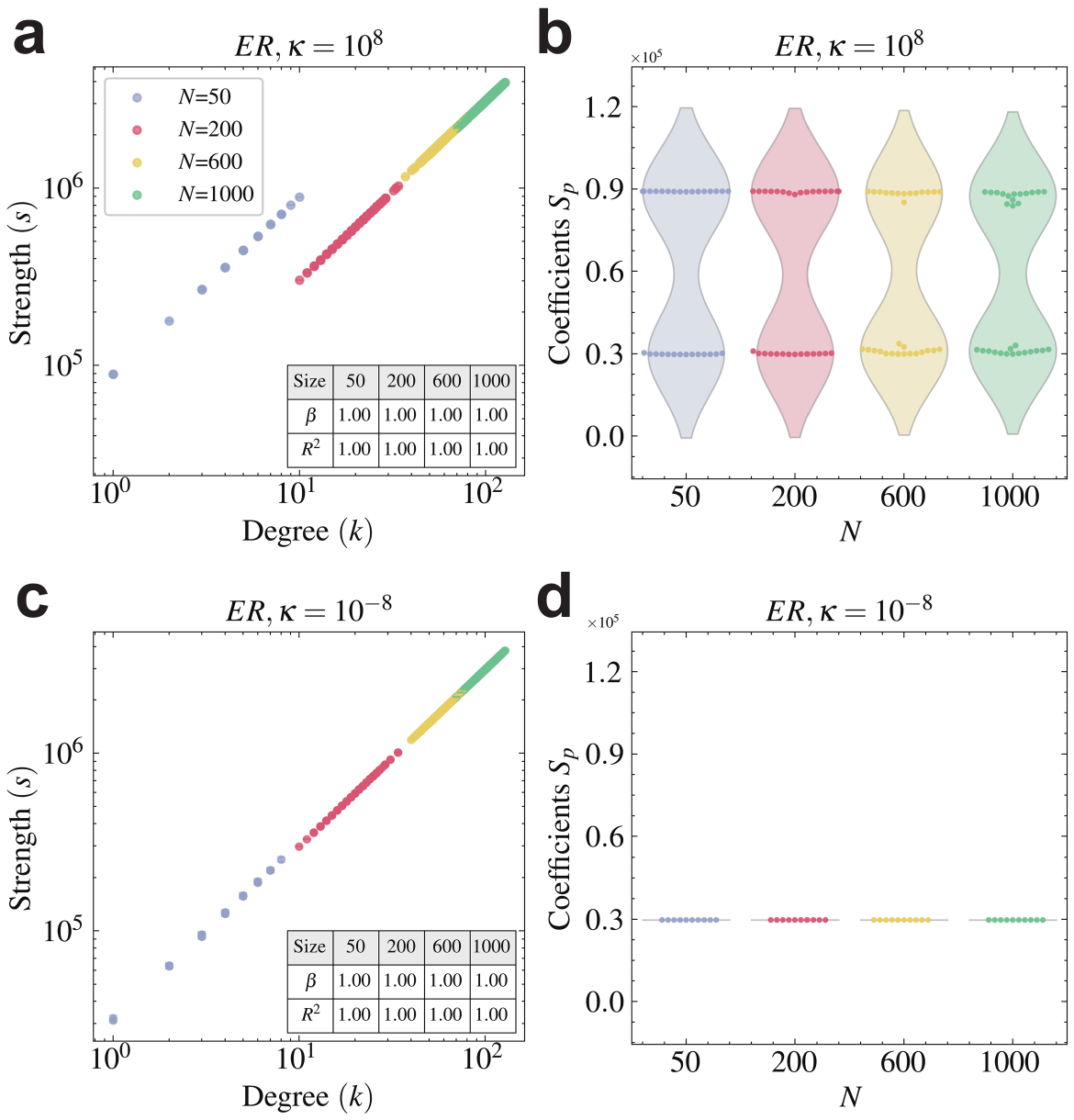}
	\caption{\textbf{Empirical validation of strength--degree linearity on ER networks of varying sizes.} 
        (a) and (c) display the relationship between node strength and degree on ER networks of different sizes ($N = 50$, $200$, $600$, $1000$), under strong ($\kappa = 10^8$) and weak ($\kappa = 10^{-8}$) selection intensities, respectively. Each point represents a node; colors distinguish network sizes. The fitted model $S \propto k^\beta$ is applied in log--log space, with fitted slopes $\beta$ and $R^2$ scores summarized in the insets. 
        (b) and (d) show the distributions of fitted linear coefficients $S_p$ (intercepts from panels (a) and (c), respectively), aggregated over 10 independent trials for each network size.
        Under strong selection, neutral drift yields stochastic convergence to either cooperation or defection in each trial, producing two parallel lines in (a) and a bimodal distribution in (b), aligned with the payoff ratio ($r:p = 3\!:\!1$). 
        Under weak selection, strategy convergence becomes deterministic (toward defection), collapsing the strength--degree relation to a single trend and eliminating bimodality, as shown in (c) and (d).
        All simulations use $r = 3$, $p = 1$, and time series length $T = 10{,}000$.}
\label{fig:linear_ER}
\end{figure*}

We focus first on ER networks. For each node, we compute the spectral strength by applying the DFT to its converged payoff time series, and compare it with its ground-truth degree. Fig~\ref{fig:linear_ER}a and~\ref{fig:linear_ER}c show the resulting log-log scatter plots under two regimes of selection intensity: high (\( \kappa = 10^8 \)) and low (\( \kappa = 10^{-8} \)). In both cases, a tight linear pattern is observed, with regression slopes \( \beta \approx 1 \) and high coefficients of determination \( R^2 \), indicating consistent scaling of the form \( S \propto k \). These results confirms the structural interpretability of the spectral strength as a linear proxy for degree across different network scales.

Under strong selection (\( \kappa = 10^8 \)), agents imitate others nearly at random (neutral drift), causing the system to stochastically converge to either full cooperation or full defection in each trial. These two absorbing states yield distinct intercepts in the strength--degree relationship, corresponding to payoff baselines \( r = 3 \) and \( p = 1 \), respectively. As shown in Fig.~\ref{fig:linear_ER}b, this trial-level bifurcation results in a bimodal distribution of the fitted linear coefficients \( S_p \), aligned with the $3\!:\!1$ payoff ratio. Fig~\ref{fig:Supple_coeff_N_b} in \nameref{Sec_SI} further confirms that reducing \( r \) to 2 yields a shift to a \( 2\!:\!1 \) intercept ratio under the same selection regime, reinforcing the payoff-induced nature of this trial-level bifurcation.

In contrast, under weak selection (\( \kappa = 10^{-8} \)), agents deterministically imitate higher-payoff neighbors, leading to global convergence toward defection—the dominant Nash equilibrium. As shown in Fig.~\ref{fig:linear_ER}c, all nodes collapse onto a single linear trend with consistent slope and intercept, indicating a structurally homogeneous response. The bimodal distribution of linear coefficients observed under neutral drift disappears entirely (Fig.~\ref{fig:linear_ER}d), confirming that behavioral convergence removes payoff-induced heterogeneity. These results further underscore the structural origin of the strength--degree linearity, which persists independently of dynamic variability across trials.

A closer comparison between different cooperation payoffs reveals that the spectral strength distributions remain nearly identical across $r = 3$ and $r = 2$ (Fig.~\ref{fig:Supple_compare_r} in \nameref{Sec_SI}). This suggests that, under weak selection, the convergence to full defection effectively suppresses stochastic variations in node-level payoffs, rendering the linear coefficients insensitive to moderate changes in the payoff parameters. The strength--degree scaling thus exhibits strong robustness, even in the presence of varying incentive structures.

These patterns persist across BA and WS networks, as demonstrated in Figs~\ref{fig:linear_BA} and~\ref{fig:linear_WS} in \nameref{Sec_SI}, confirming the universality of the strength--degree linearity across diverse topologies. Taken together, these results validate the theoretical foundation of our framework: the spectral strength derived from frequency-domain analysis serves as a reliable linear proxy for node degree, robust to variations in network topology, selection intensity, and payoff parameters. This structurally grounded and dynamically consistent relationship provides a principled basis for inferring global network structure from local behavioral dynamics.

\subsection{Network reconstruction}

Having established the theoretical validity of the strength–degree linear relationship, we now assess the effectiveness of our DFT-based framework for reconstructing network topology directly from node-level behavioral dynamics. The input to this method is solely the node-level payoff time series, and the output is the inferred adjacency matrix of the network.

\begin{figure*}[htbp]
	\centering
        \includegraphics[width=\textwidth]{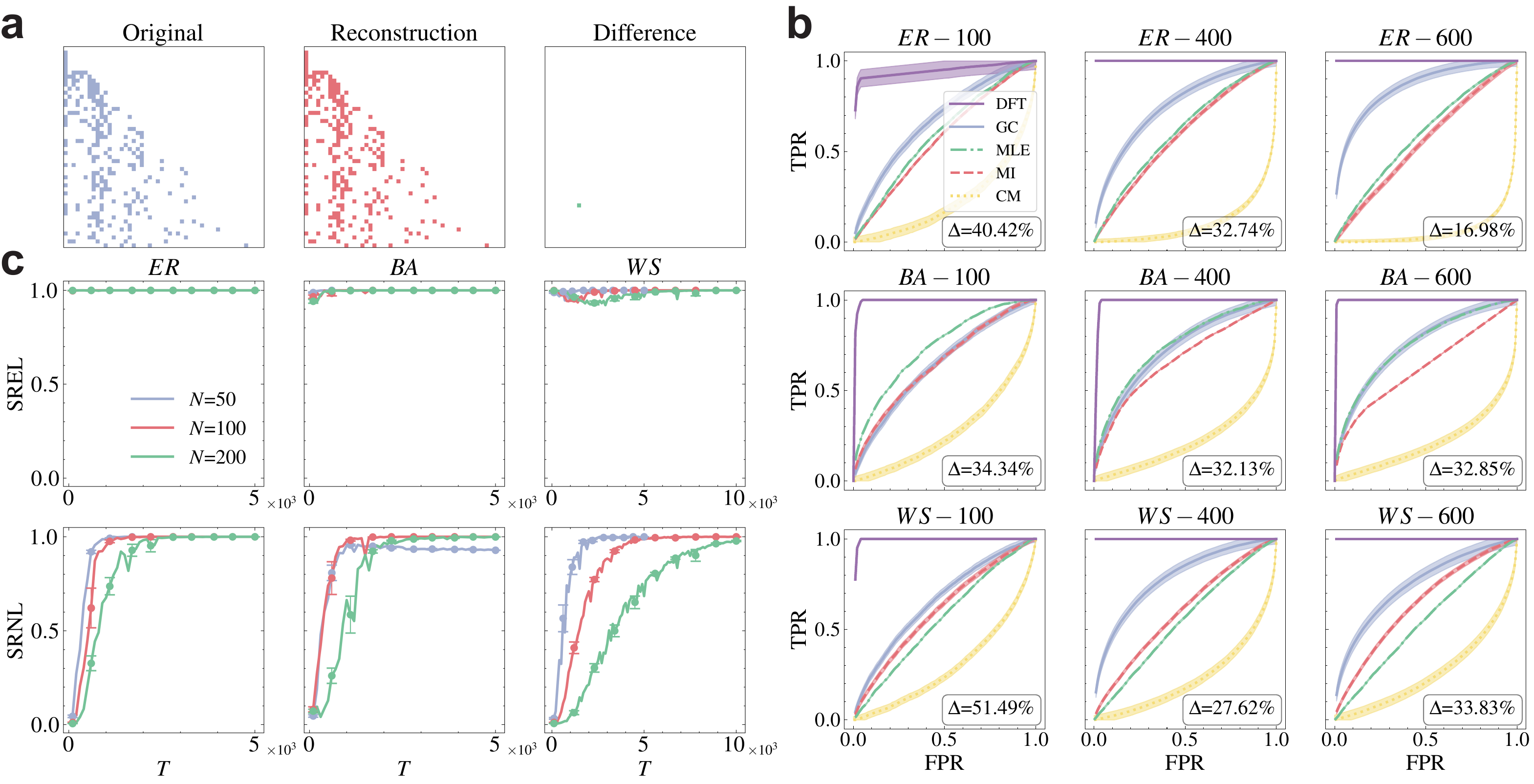}
         \caption{\textbf{Performance evaluation of network reconstruction using the DFT method.} 
         (a) Visualization of reconstruction results on a BA network with 50 nodes. The \textit{Original Matrix} shows ground-truth links (colored entries), while the \textit{Reconstruction Matrix} marks predicted links. The \textit{Difference Matrix} highlights mismatches, with green entries denoting errors.
         (b) ROC curves comparing the DFT method against four baselines (GC, CM, MLE, MI) across ER, BA, and WS networks of varying sizes. The area under the curve (AUC) is used as the evaluation metric, and the relative improvement is computed as \( \Delta = (\text{AUC}_{\text{DFT}} - \text{AUC}_{\text{best baseline}})/\text{AUC}_{\text{best baseline}} \). 
         (c) Impact of time series length on reconstruction accuracy, evaluated using SREL and SRNL metrics. 
         Results in (b) and (c) are averaged over 10 independent trials. Synthetic network properties are provided in Table~\ref{tab:density_nets}}.
    \label{fig:reconstruction_results}
\end{figure*}

We first evaluate reconstruction accuracy on a small synthetic BA network (50 nodes, 225 existing edges, 1,000 non-existing edges). Fig~\ref{fig:reconstruction_results}a visualizes the reconstructed adjacency matrix compared with the ground-truth. The difference matrix reveals nearly perfect reconstruction (SREL = 1.000, SRNL = 0.999), indicating only one false-positive edge. This preliminary result confirms the capability of the DFT approach to precisely recover structural information from limited dynamical observations.

Next, we systematically evaluate the performance on larger synthetic networks. Fig~\ref{fig:reconstruction_results}b compares the receiver operating characteristic (ROC) curves of our method with widely-used baseline methods (GC, CM, MLE, MI) across ER, BA, and WS networks of varying sizes. Each ROC curve is generated by varying the threshold for predicted connection strengths, yielding true-positive and false-positive rates across different binary classification scenarios. Our DFT method consistently achieves the highest area under the curve (AUC) values, substantially outperforming baseline methods, indicating superior performance in both sensitivity and specificity across diverse networks. The maximum relative improvement of DFT over the best baseline method reaches 51.49\%.

\begin{table*}[htbp]
\caption{\textbf{Performance comparison of network reconstruction on synthetic networks.}
This table reports the reconstruction performance of the proposed DFT-based method against four baseline methods—GC, CM, MLE, and MI—across BA, ER, and WS networks of varying sizes. Each score represents the mean \(\pm\) standard deviation over 10 trials, with time series length set to \(T=20N\). Boldface indicates the best-performing method. Synthetic network properties are listed in Table~\ref{tab:density_nets}.}
\label{tab:srenl_syn}
\tiny
\centering\small
\setlength{\tabcolsep}{2pt}
\begin{tabular*}{\textwidth}{@{\extracolsep\fill}l l c c c c c}
\toprule
Network        & Metric & GC                 & CM                 & MLE                & MI                 & DFT                  \\
\midrule
\multirow{2}{*}{BA-100}
               & SREl      & \(0.639 \pm 0.042\) & \(0.296 \pm 0.151\) & \(0.565 \pm 0.016\) & \(0.411 \pm 0.070\) & \(\boldsymbol{1.000} \pm 0.000\) \\
               & SRNl      & \(0.555 \pm 0.033\) & \(0.705 \pm 0.150\) & \(0.674 \pm 0.023\) & \(0.689 \pm 0.063\) & \(\boldsymbol{0.995} \pm 0.002\) \\
\addlinespace
\multirow{2}{*}{BA-200}
               & SREL      & \(0.609 \pm 0.018\) & \(0.573 \pm 0.158\) & \(0.694 \pm 0.013\) & \(0.610 \pm 0.059\) & \(\boldsymbol{1.000} \pm 0.000\) \\
               & SRNL      & \(0.617 \pm 0.019\) & \(0.428 \pm 0.158\) & \(0.549 \pm 0.019\) & \(0.528 \pm 0.066\) & \(\boldsymbol{0.998} \pm 0.000\) \\
\addlinespace
\multirow{2}{*}{BA-400}
               & SREL      & \(0.673 \pm 0.038\) & \(0.900 \pm 0.100\) & \(0.548 \pm 0.012\) & \(0.462 \pm 0.073\) & \(\boldsymbol{1.000} \pm 0.000\) \\
               & SRNL      & \(0.683 \pm 0.021\) & \(0.100 \pm 0.100\) & \(0.708 \pm 0.015\) & \(0.650 \pm 0.075\) & \(\boldsymbol{0.976} \pm 0.003\) \\
\addlinespace
\multirow{2}{*}{BA-600}
               & SREL      & \(0.749 \pm 0.019\) & \(0.800 \pm 0.133\) & \(0.588 \pm 0.007\) & \(0.246 \pm 0.009\) & \(\boldsymbol{1.000} \pm 0.000\) \\
               & SRNL      & \(0.713 \pm 0.025\) & \(0.200 \pm 0.133\) & \(0.676 \pm 0.010\) & \(0.845 \pm 0.008\) & \(\boldsymbol{0.997} \pm 0.001\) \\
\addlinespace
\multirow{2}{*}{ER-100}
               & SREL      & \(0.635 \pm 0.039\) & \(0.606 \pm 0.160\) & \(0.694 \pm 0.023\) & \(0.499 \pm 0.036\) & \(\boldsymbol{1.000} \pm 0.000\) \\
               & SRNL      & \(0.609 \pm 0.032\) & \(0.397 \pm 0.161\) & \(0.367 \pm 0.023\) & \(0.639 \pm 0.035\) & \(\boldsymbol{0.996} \pm 0.003\) \\
\addlinespace
\multirow{2}{*}{ER-200}
               & SREL      & \(0.678 \pm 0.025\) & \(0.800 \pm 0.133\) & \(0.687 \pm 0.032\) & \(0.629 \pm 0.053\) & \(\boldsymbol{1.000} \pm 0.000\) \\
               & SRNL      & \(0.606 \pm 0.037\) & \(0.201 \pm 0.133\) & \(0.413 \pm 0.033\) & \(0.451 \pm 0.060\) & \(\boldsymbol{1.000} \pm 0.000\) \\
\addlinespace
\multirow{2}{*}{ER-400}
               & SREL      & \(0.718 \pm 0.022\) & \(0.700 \pm 0.153\) & \(0.527 \pm 0.027\) & \(0.681 \pm 0.045\) & \(\boldsymbol{1.000} \pm 0.000\) \\
               & SRNL      & \(0.660 \pm 0.024\) & \(0.300 \pm 0.153\) & \(0.569 \pm 0.029\) & \(0.373 \pm 0.052\) & \(\boldsymbol{1.000} \pm 0.000\) \\
\addlinespace
\multirow{2}{*}{ER-600}
               & SREL      & \(0.757 \pm 0.018\) & \(0.600 \pm 0.163\) & \(0.536 \pm 0.008\) & \(0.673 \pm 0.026\) & \(\boldsymbol{1.000} \pm 0.000\) \\
               & SRNL      & \(0.749 \pm 0.020\) & \(0.400 \pm 0.163\) & \(0.552 \pm 0.009\) & \(0.398 \pm 0.033\) & \(\boldsymbol{1.000} \pm 0.000\) \\
\addlinespace
\multirow{2}{*}{WS-100}
               & SREL      & \(0.585 \pm 0.040\) & \(0.606 \pm 0.160\) & \(0.694 \pm 0.023\) & \(0.499 \pm 0.036\) & \(\boldsymbol{1.000} \pm 0.000\) \\
               & SRNL      & \(0.554 \pm 0.036\) & \(0.397 \pm 0.161\) & \(0.367 \pm 0.023\) & \(0.639 \pm 0.035\) & \(\boldsymbol{0.996} \pm 0.003\) \\
\addlinespace
\multirow{2}{*}{WS-200}
               & SREL      & \(0.633 \pm 0.023\) & \(0.800 \pm 0.133\) & \(0.473 \pm 0.032\) & \(0.618 \pm 0.028\) & \(\boldsymbol{1.000} \pm 0.000\) \\
               & SRNL      & \(0.644 \pm 0.024\) & \(0.201 \pm 0.133\) & \(0.574 \pm 0.031\) & \(0.537 \pm 0.037\) & \(\boldsymbol{0.995} \pm 0.002\) \\
\addlinespace
\multirow{2}{*}{WS-400}
               & SREL      & \(0.695 \pm 0.040\) & \(0.700 \pm 0.153\) & \(0.539 \pm 0.016\) & \(0.553 \pm 0.045\) & \(\boldsymbol{1.000} \pm 0.000\) \\
               & SRNL      & \(0.728 \pm 0.025\) & \(0.300 \pm 0.153\) & \(0.515 \pm 0.017\) & \(0.592 \pm 0.037\) & \(\boldsymbol{0.999} \pm 0.001\) \\
\addlinespace
\multirow{2}{*}{WS-600}
               & SREL      & \(0.641 \pm 0.034\) & \(0.500 \pm 0.167\) & \(0.529 \pm 0.025\) & \(0.620 \pm 0.041\) & \(\boldsymbol{1.000} \pm 0.000\) \\
               & SRNL      & \(0.706 \pm 0.019\) & \(0.500 \pm 0.167\) & \(0.518 \pm 0.025\) & \(0.607 \pm 0.044\) & \(\boldsymbol{1.000} \pm 0.000\) \\
\addlinespace
\multirow{2}{*}{Average}
               & SREL      & \(0.668\)           & \(0.657\)           & \(0.590\)           & \(0.542\)           & \(\boldsymbol{1.000}\)           \\
               & SRNL      & \(0.652\)           & \(0.344\)           & \(0.540\)           & \(0.579\)           & \(\boldsymbol{0.996}\)           \\
\bottomrule
\end{tabular*}
\end{table*}

Quantitative performance metrics—SREL and SRNL—are reported in Table~\ref{tab:srenl_syn} to assess reconstruction quality. Both metrics range from 0 to 1, and higher values indicate more accurate recovery of true and false links, respectively. The proposed DFT-based method achieves superior performance across all tested scenarios. Specifically, SREL reaches a perfect score of 1.000 in every case, indicating that all actual edges are correctly identified. SRNL scores are similarly high, consistently approaching unity. Compared to the best-performing baselines, our method improves the average SREL and SRNL by 50\% and 53\%, respectively, with maximum gains of 76.99\% (BA-100, SREL) and 81.79\% (BA-200, SRNL).

To evaluate practical applicability, we extend our experiments to empirical networks. Results presented in Table~\ref{tab:srenl_emp} reveal that the DFT method achieves the highest SREL across all empirical cases, markedly surpassing baseline methods. The largest improvements are observed in the Madrid network (97\% SREL increase) and the USAir network (41\% SRNL increase). Although the DFT method's SRNL scores on empirical data are slightly lower than those on synthetic data—reflecting greater challenges in accurately excluding non-existing edges in real-world data—its overall performance remains clearly superior. This difference may be attributed to insufficient length of empirical time series, highlighting the need for longer observational data to effectively discriminate absent interactions.

\begin{table*}[htbp]
\caption{\textbf{Performance comparison of network reconstruction on empirical networks.}
We evaluate the DFT-based method and four baselines (GC, CM, MLE, MI) on 11 real-world networks. Each score represents the mean \(\pm\) standard deviation over 10 trials, with time series length set to \(T=20N\). Boldface highlights the best performance for each network. Network details are provided in Table~\ref{tab:table1}.}
\label{tab:srenl_emp}
\centering\small
\setlength{\tabcolsep}{2pt}
\begin{tabular*}{\textwidth}{@{\extracolsep\fill}l l c c c c c}
\toprule
Network   & Metric & GC                & CM                & MLE               & MI                & DFT                \\
\midrule
\multirow{2}{*}{Sheep}
          & SREL   & \(0.583 \pm 0.054\) & \(0.332 \pm 0.124\) & \(0.700 \pm 0.008\) & \(0.697 \pm 0.032\) & \(\boldsymbol{1.000} \pm 0.000\) \\
          & SRNL   & \(0.492 \pm 0.054\) & \(\boldsymbol{0.706} \pm 0.125\) & \(0.657 \pm 0.005\) & \(0.604 \pm 0.035\) & \(0.634 \pm 0.039\) \\
\addlinespace
\multirow{2}{*}{Dolphin}
          & SREL   & \(0.738 \pm 0.042\) & \(0.615 \pm 0.095\) & \(0.763 \pm 0.058\) & \(0.516 \pm 0.041\) & \(\boldsymbol{1.000} \pm 0.000\) \\
          & SRNL   & \(0.724 \pm 0.034\) & \(0.548 \pm 0.099\) & \(0.481 \pm 0.051\) & \(\boldsymbol{0.772} \pm 0.025\) & \(0.758 \pm 0.079\) \\
\addlinespace
\multirow{2}{*}{Madrid}
          & SREL   & \(0.567 \pm 0.053\) & \(0.411 \pm 0.085\) & \(0.729 \pm 0.066\) & \(0.508 \pm 0.029\) & \(\boldsymbol{1.000} \pm 0.000\) \\
          & SRNL   & \(0.707 \pm 0.036\) & \(0.712 \pm 0.079\) & \(0.544 \pm 0.059\) & \(0.778 \pm 0.026\) & \(\boldsymbol{0.963} \pm 0.008\) \\
\addlinespace
\multirow{2}{*}{Sandi}
          & SREL   & \(0.794 \pm 0.032\) & \(0.580 \pm 0.050\) & \(0.621 \pm 0.009\) & \(0.528 \pm 0.035\) & \(\boldsymbol{0.999} \pm 0.001\) \\
          & SRNL   & \(\boldsymbol{0.803} \pm 0.032\) & \(0.767 \pm 0.035\) & \(0.701 \pm 0.021\) & \(0.790 \pm 0.035\) & \(0.767 \pm 0.017\) \\
\addlinespace
\multirow{2}{*}{Retweet}
          & SREL   & \(0.857 \pm 0.029\) & \(0.591 \pm 0.060\) & \(0.820 \pm 0.017\) & \(0.339 \pm 0.059\) & \(\boldsymbol{1.000} \pm 0.000\) \\
          & SRNL   & \(0.795 \pm 0.051\) & \(0.702 \pm 0.054\) & \(0.581 \pm 0.021\) & \(0.815 \pm 0.048\) & \(\boldsymbol{0.975} \pm 0.003\) \\
\addlinespace
\multirow{2}{*}{Songbird}
          & SREL   & \(0.607 \pm 0.019\) & \(0.614 \pm 0.099\) & \(0.594 \pm 0.010\) & \(0.544 \pm 0.027\) & \(\boldsymbol{0.996} \pm 0.002\) \\
          & SRNL   & \(0.566 \pm 0.013\) & \(0.465 \pm 0.104\) & \(0.679 \pm 0.011\) & \(\boldsymbol{0.707} \pm 0.029\) & \(0.646 \pm 0.018\) \\
\addlinespace
\multirow{2}{*}{Email}
          & SREL   & \(0.709 \pm 0.025\) & \(0.377 \pm 0.099\) & \(0.526 \pm 0.014\) & \(0.519 \pm 0.024\) & \(\boldsymbol{1.000} \pm 0.000\) \\
          & SRNL   & \(0.708 \pm 0.021\) & \(0.698 \pm 0.103\) & \(0.695 \pm 0.015\) & \(0.684 \pm 0.024\) & \(\boldsymbol{0.839} \pm 0.088\) \\
\addlinespace
\multirow{2}{*}{Mouse}
          & SREL   & \(0.871 \pm 0.029\) & \(0.723 \pm 0.044\) & \(0.983 \pm 0.001\) & \(0.041 \pm 0.011\) & \(\boldsymbol{1.000} \pm 0.000\) \\
          & SRNL   & \(0.782 \pm 0.058\) & \(0.715 \pm 0.047\) & \(0.908 \pm 0.002\) & \(0.994 \pm 0.001\) & \(\boldsymbol{0.985} \pm 0.003\) \\
\addlinespace
\multirow{2}{*}{USAir}
          & SREL   & \(0.701 \pm 0.014\) & \(0.347 \pm 0.050\) & \(0.735 \pm 0.011\) & \(0.206 \pm 0.010\) & \(\boldsymbol{1.000} \pm 0.000\) \\
          & SRNL   & \(0.686 \pm 0.051\) & \(0.773 \pm 0.043\) & \(0.660 \pm 0.015\) & \(0.957 \pm 0.004\) & \(\boldsymbol{0.932} \pm 0.017\) \\
\addlinespace
\multirow{2}{*}{Twitter}
          & SREL   & \(0.921 \pm 0.017\) & \(0.590 \pm 0.043\) & \(0.864 \pm 0.011\) & \(0.107 \pm 0.006\) & \(\boldsymbol{1.000} \pm 0.000\) \\
          & SRNL   & \(0.933 \pm 0.014\) & \(0.823 \pm 0.018\) & \(0.579 \pm 0.016\) & \(0.962 \pm 0.003\) & \(\boldsymbol{0.956} \pm 0.002\) \\
\addlinespace
\multirow{2}{*}{Crime}
          & SREL   & \(0.925 \pm 0.024\) & \(0.212 \pm 0.097\) & \(0.691 \pm 0.014\) & \(0.366 \pm 0.019\) & \(\boldsymbol{1.000} \pm 0.000\) \\
          & SRNL   & \(0.939 \pm 0.019\) & \(0.828 \pm 0.092\) & \(0.594 \pm 0.006\) & \(0.818 \pm 0.024\) & \(\boldsymbol{0.999} \pm 0.000\) \\
\midrule
\multirow{2}{*}{Average}
          & SREL   & \(0.752\) & \(0.490\) & \(0.730\) & \(0.397\) & \(\boldsymbol{1.000}\) \\
          & SRNL   & \(0.740\) & \(0.703\) & \(0.644\) & \(0.807\) & \(\boldsymbol{0.859}\) \\
\botrule
\end{tabular*}
\end{table*}

Beyond accuracy, a key advantage of the DFT-based approach is its scalability. While most baseline methods are computationally limited to networks with only a few hundred nodes, our method consistently maintains high reconstruction accuracy on networks approaching a thousand nodes. This scalability is essential for real-world applications, where large-scale structures often exceed the capacity of conventional techniques. By leveraging efficient spectral analysis, the DFT method achieves a favorable balance between performance and computational efficiency, enabling accurate inference on both synthetic and empirical networks at scale.

To further explore the relationship between inference accuracy and observational data length, we analyze the DFT reconstruction performance across varying time series lengths (Fig~\ref{fig:reconstruction_results}c). Results demonstrate that reconstruction accuracy generally improves with longer time series. Importantly, SREL converges rapidly to near-perfect values even with relatively short sequences, whereas SRNL converges more gradually, especially in larger networks. Among network types, ER networks exhibit the fastest convergence, followed by BA and then WS networks. The delayed convergence in WS networks likely arises from their pronounced small-world structure and relatively uniform node degrees, which necessitate more extended temporal observations to discern subtle topological features. These findings underscore the importance of adequate temporal data length for the effective application of frequency-domain network inference.

Collectively, these analyses confirm the robustness, scalability, and practical applicability of our DFT-based inference framework, demonstrating its substantial advantages over existing methods across diverse synthetic and empirical network structures.

\subsection{Detect hidden nodes and edges}

Building on the accurate reconstruction of observable network structure, our DFT-based method also enables reliable identification of hidden nodes and their latent connections. Specifically, we consider two distinct scenarios: one involving a single known hidden node, and another involving multiple potentially unknown hidden nodes.

In the single hidden node scenario, we assume that one hidden node exists in the system, which is unobservable and cannot be perturbed. Only time series from the observable nodes are available; no prior topological information is assumed. The DFT-based method enables both the reconstruction of the network and the localization of the hidden node by identifying all observable nodes that connect to it (i.e., the external or hidden edges).

\begin{figure*}[htbp]
	\centering
        \includegraphics[width=.7\textwidth]{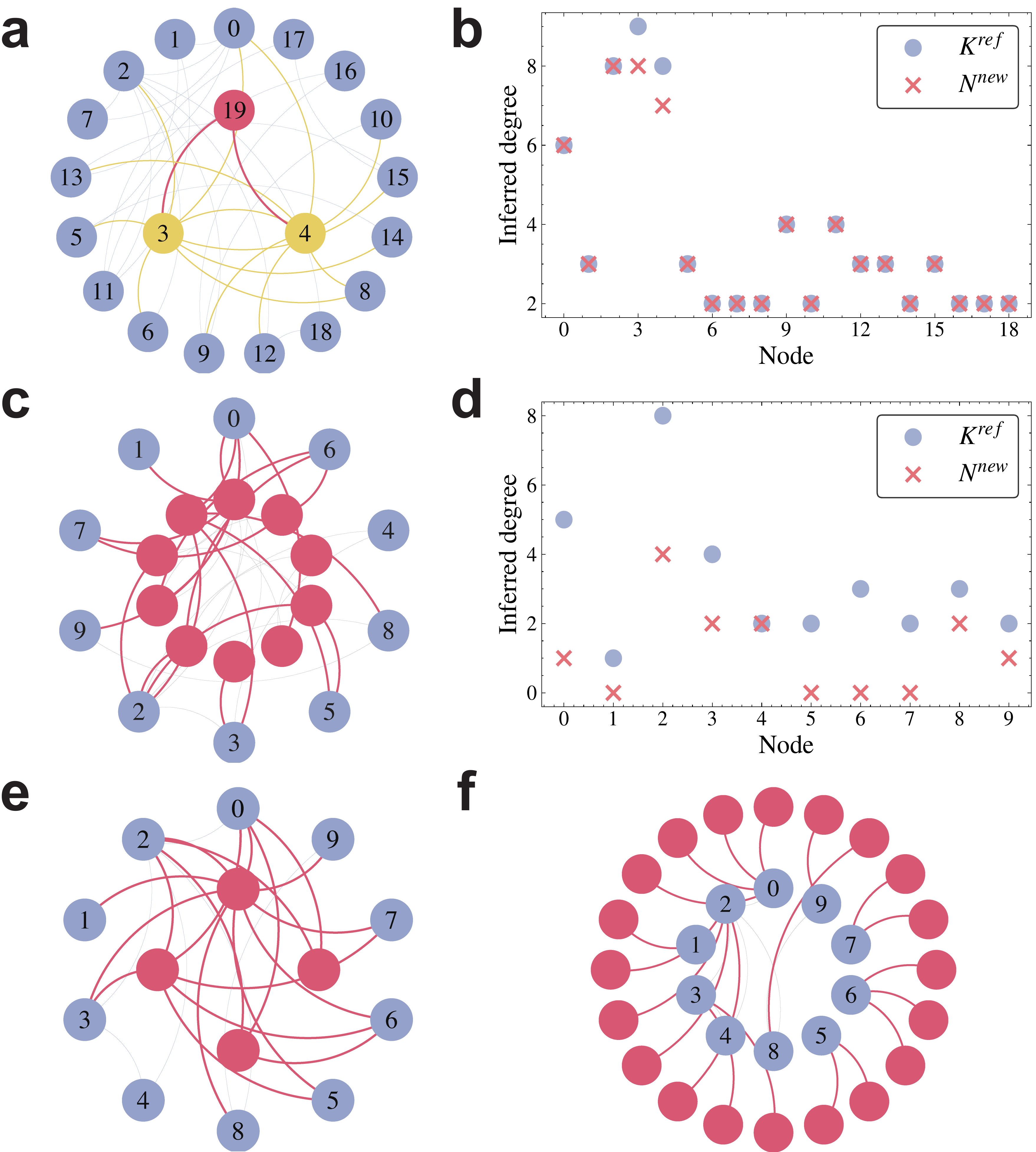}
    \caption{\textbf{Identification of hidden nodes and their external connections.} 
    (a,b) Single hidden node scenario. (a) Ground-truth network with one hidden node (red), observable neighbors (blue), and unaffected nodes (gray). Red edges indicate external edges. (b) Comparison of inferred degrees before intervention ($K^{\mathrm{ref}}$) and the number of identified neighbors after intervention ($N^{\mathrm{new}}$). Nodes 3 and 4 each exhibit a unit discrepancy, suggesting a shared hidden neighbor. 
    (c,d) Multiple hidden nodes scenario. (c) Ground-truth network with multiple hidden nodes and observable neighbors. External connections (red edges) are successfully identified by perturbation analysis. (d) Comparison between $K^{\mathrm{ref}}$ and $N^{\mathrm{new}}$ across all observable nodes reveals external links. 
    (e,f) Extreme realizations of hidden node estimation based on (d). (e) Lower-bound configuration where the number of hidden nodes equals the maximum number of external links among all observable neighbors—i.e., observable nodes share hidden nodes as much as possible. (f) Upper-bound configuration where each external edge originates from a distinct hidden node. 
    All results are obtained with time-series length $T=400$.}
    \label{fig:hidden_node_results}
\end{figure*}

Fig~\ref{fig:hidden_node_results}a illustrates this process using a simple BA network where node 19 is hidden. We begin by collecting payoff time series from all observable nodes and compute their inferred degrees using the DFT method. Next, we iteratively suppress each observable node’s dynamics and identify its neighborhood from the resulting time series. As shown in Fig~\ref{fig:hidden_node_results}b, for nodes 3 and 4, we find that their inferred degrees before intervention exceed the number of post-intervention neighbors by exactly one, indicating that each has an unobservable neighbor. This indicates a hidden node directly connected to both nodes 3 and 4.

Table~\ref{tab:hidden_nodes_summary} summarizes the performance of the DFT method in detecting a single hidden node across both empirical and synthetic networks. To evaluate accuracy, we define the metric \textbf{Accuracy-I} as the ratio between the number of correctly identified neighbors of the hidden node and its true number of neighbors. To enable an internal comparison of identification consistency across known and variable settings, we fix the hidden node across all trials for empirical networks, while randomly selecting it in each trial for synthetic networks. In all cases, the method achieves high accuracy: five networks reach perfect identification, and all others exceed 0.9. Moreover, accuracy remains robust as network size increases, confirming the method’s scalability and reliability.

\begin{table}[htbp]
\centering
\caption{Summary of single hidden node identification across empirical and synthetic networks. In empirical settings, the hidden node is fixed across trials; in synthetic settings, it is randomly selected in each run. As a result, the number of neighbors varies across trials and is marked with an asterisk (*). Accuracy is reported as the mean ± standard deviation over 50 trials, with time series length set to \(T=20N\). Network details are provided in Tables~\ref{tab:table1} and~\ref{tab:density_nets}.}
\label{tab:hidden_nodes_summary}
\begin{tabular}{lccc}
\toprule
Network                  & Hidden node index & Number of neighbors of hidden node & Accuracy-I              \\
\midrule
Sheep                    & 17                & 16                                 & $ 0.960 \pm 0.027$      \\
Songbird                 & 54                & 11                                 & \(1.000 \pm 0.000\)     \\
Madrid                   & 30                & 6                                  & \(0.997 \pm 0.003\)      \\
Email                    & 96                & 2                                  & \(1.000 \pm 0.000\)      \\
ER-200                   &random             & *                                  & \(1.000 \pm 0.000\)      \\
ER-600                   &random             & *                                  & \(0.920 \pm 0.071\)      \\
BA-200                   &random             & *                                  & $ 0.957 \pm 0.013$      \\
BA-600                   &random             & *                                  & $ 1.000 \pm 0.000$      \\
WS-200                   &random             & *                                  & $ 0.939 \pm 0.019$      \\
WS-600                   &random             & *                                  & $ 0.917 \pm 0.017$      \\

\bottomrule
\end{tabular}
\end{table}

In the scenario with multiple hidden nodes, neither their existence nor their number is known, and only dynamical time series from observable nodes are available. The DFT method enables network reconstruction while simultaneously identifying all external edges through inconsistencies between inferred degrees and the number of neighbors identified after intervention. If, for all observable nodes, the inferred degrees match the number of post-intervention neighbors, no hidden nodes are present. Otherwise, the difference between a node’s inferred degree and its post-intervention neighborhood size indicates the number of external edges linking that node to hidden nodes. The outputs of this procedure include the exact number and location of all external edges (i.e., their observable endpoints), which allows estimation of a bounded range for the number of hidden nodes.

Fig~\ref{fig:hidden_node_results}c shows a sample network with 10 observable and 10 hidden nodes. Red edges indicate external connections. The DFT method detects these links by comparing each node’s inferred degree (before intervention) with the number of neighbors identified after intervention. As shown in Fig~\ref{fig:hidden_node_results}d, node 1 exhibits a surplus of 4 in its inferred degree, corresponding precisely to four hidden connections. Nodes with no external links, such as node 8, show an exact match between pre- and post-intervention values. Applying this procedure to all observable nodes enables precise identification of all external edges.

Table~\ref{tab:external_edges} summarizes the performance of the method across networks with different numbers of hidden nodes. To evaluate accuracy, we define \textbf{Accuracy-II} as the ratio between the number of correctly identified external edges and the true number of such edges. The results show that the method consistently achieves high accuracy, with all cases exceeding 0.92.

\begin{table}[htbp]
\centering
\tiny
\caption{Summary of external edge identification under varying hidden node and external edge settings in the BA-600 network (see Table~\ref{tab:density_nets}). Accuracy-II is reported as the mean ± standard deviation over 10 trials, with time series length set to \(T=20N\). Bounds indicate the estimated lower and upper limits on the number of hidden nodes.}
\label{tab:external_edges}
\begin{tabular}{lcccc}
\toprule
Number of hidden nodes     & Ave. number of external edges & Ave. number of neighbors & Accuracy-II            & Bounds    \\
\midrule
10                         & 79                            & 74                       & $0.973\pm0.003$     & [2,79]    \\  
20                         & 92                            & 82                       & $0.951\pm0.021$     & [3,90]    \\  
40                         & 117                           & 94                       & $0.922\pm0.017$     & [5,110]   \\  
60                         & 213                           & 165                      & $0.986\pm0.013$     & [7,213]   \\  
80                         & 313                           & 225                      & $0.972\pm0.021$     & [6,310]   \\

\bottomrule
\end{tabular}
\end{table}

Based on the identified external edges, we estimate a feasible range for the number of hidden nodes. The minimum and maximum estimates correspond to two limiting configurations: one where observable nodes maximally share hidden neighbors, and another where each external edge originates from a distinct hidden node. These two cases are illustrated in Figs~\ref{fig:hidden_node_results}e and~\ref{fig:hidden_node_results}f, with 4 and 20 hidden nodes, respectively. Detailed results under varying hidden node and external edge settings in the BA-600 network are provided in Table~\ref{tab:external_edges}.

As hidden nodes are unobservable and cannot be directly perturbed, it is not possible to determine their exact neighbors or internal connections. While precise identification remains elusive, the DFT-based method offers a tractable and scalable solution by inferring all external links and bounding the number of hidden nodes.

\section{Methods}
\label{sec:methods}
This section introduces the core methodology developed in this study. At the heart of our approach lies a novel frequency-domain framework for inferring complex network structure based on node-level payoff dynamics under evolutionary games. We begin by extracting meaningful frequency-domain features from time series data using the DFT. Building on the empirical discovery of a robust linear relationship between spectral strength and node degree, we design a reconstruction algorithm that accurately infers network topology. Furthermore, by selectively perturbing node participation, we extend the framework to identify hidden nodes and their external edges. These methodological advances form the methodological foundation of our work.

\subsection{Frequency-domain Feature Extraction via DFT}

The DFT is a classical method for analyzing the frequency components of discrete-time signals. For a discrete sequence of length $T$, denoted as ${x[t]}_{t=0}^{T-1}$, its DFT is defined as:
\begin{equation}
\hat{x}[f] = \sum_{t=0}^{T-1} x[t] e^{-i \frac{2\pi}{T}ft},
\end{equation}
where \(e\) represents the base of the natural logarithm, $f = 0, 1, ..., T-1$, $i$ is the imaginary unit, and $\hat{x}[f]$ represents the complex Fourier coefficient at frequency $f/T$.

In this study, we focus on real-valued payoff sequences $P_i(t)$ associated with each node $i$, representing cumulative rewards accrued during repeated interactions. These real signals exhibit conjugate symmetry in their DFT, meaning $\hat{P}_i[f] = \overline{\hat{P}_i[T-f]}$, and thus the magnitude spectrum $|\hat{P}_i[f]|$ suffices to characterize the frequency domain information. 

Moreover, it has been empirically confirmed that the phase information and the DC component ($f = 0$) do not significantly contribute to the inference task. Therefore, we retain only the magnitudes of non-zero frequency components: $|\hat{P}_i[f]|$, $f \in F = {1, 2, ..., T/2}$, which serve as input features for subsequent inference procedures. This reduction improves computational efficiency and model scalability.

\subsection{Linear Relationship Between Node Dynamics and Degree}

Let $G = (V, E)$ denote a network with node set $V$ and edge set $E$. In each game round, every node $i \in V$ interacts with neighbors $j \in \mathcal{N}_i$ under the Prisoner’s Dilemma rules. The strategy $s_i(t) \in {0,1}$ indicates cooperation ($0$) or defection ($1$), and the payoff from interacting with $j$ is:

\begin{equation}
\pi(s_i, s_j) = r(1-s_i)(1-s_j) + s(1-s_i)s_j + t s_i(1-s_j) + p s_i s_j,
\end{equation}

where $r$, $s$, $t$, and $p$ are standard PDG parameters.

The total payoff of node $i$ at time $t$ is:

\begin{equation}
P_i(t) = \sum_{j \in \mathcal{N}_i} \pi(s_i(t), s_j(t)).
\end{equation}

This payoff can be expressed as:

\begin{equation}
P_i(t) = k_i \cdot p_i(t),
\end{equation}

where $k_i = |\mathcal{N}_i|$ is the degree of node $i$, and $p_i(t)$ is the average payoff per neighbor.

Taking the DFT of $P_i(t)$ yields:

\begin{equation}
\hat{P}_i(f) = \sum_{t=0}^{T-1} P_i(t) e^{-2\pi i f t/T} = k_i \cdot \hat{p}_i(f),
\end{equation}

implying:

\begin{equation}
|\hat{P}_i(f)| = k_i \cdot |\hat{p}_i(f)|.
\end{equation}

This establishes a direct proportional relationship between the magnitude spectrum of the payoff sequence and the node's degree, i.e., $|\hat{P}_i(f)| \propto k_i$. This linear relationship forms the theoretical foundation of our inference framework and rests upon two key assumptions: (i) the average payoff per neighbor $p_i(t)$ is statistically independent of the node's degree $k_i$; and (ii) all nodes adopt identical stochastic update rules—such as the Fermi mechanism—ensuring consistent behavioral dynamics across the network.

These assumptions are well-justified in homogeneous or moderately heterogeneous networks. Each interaction generates a payoff independent of the node’s degree, as the degree only determines the number of such interactions, not their individual outcomes. Moreover, uniform update mechanisms further decouple individual strategy dynamics from local topological features, ensuring consistent temporal patterns in $p_i(t)$ across nodes.

Nonetheless, in networks with strong heterogeneity, degree assortativity, or modular organization, these assumptions may be partially violated. Local structural effects can introduce node-dependent variations in $p_i(t)$, weakening the linearity between $|\hat{P}_i(f)|$ and $k_i$ at individual frequencies.

To mitigate deviations arising from structural heterogeneity, we aggregate frequency components to define the spectral strength of node $i$:

\begin{equation}
S_i = \frac{1}{T} \sum_{f \in F} |\hat{P}_i(f)|,
\end{equation}

where $F$ is a designated frequency set. Substituting the previous relation gives:

\begin{equation}
S_i = k_i \cdot \left( \frac{1}{T} \sum_{f \in F} |\hat{p}_i(f)| \right) = k_i \cdot S_p,
\end{equation}

where $S_p$ denotes the average spectral amplitude of $p_i(t)$ across $F$.

If $S_p$ is approximately constant across nodes—despite mild heterogeneity—then:

\begin{equation}
S_i \propto k_i.
\end{equation}

This aggregated measure reduces sensitivity to noise and local fluctuations, thereby enabling more robust and scalable inference of node degree from dynamic behavioral data, even in structurally diverse networks.

\subsection{Inferring Network Structure from Degree–Dynamics Linearity}

Building upon the established linear relationship between node strength and degree ($S_i \propto k_i$), we propose a unified, model-free framework for inferring network structure directly from the dynamical behavior of observable nodes. This framework remains effective even when the total number of nodes is unknown, thereby accommodating settings with latent or hidden components.

Our method proceeds in two main stages. First, we estimate the degree of each observable node by analyzing its payoff time series in the frequency domain. Second, we sequentially perturb each node—temporarily suppressing its activity—and observe the spectral changes in other nodes to infer local connectivity. When the observed responses cannot fully account for a node’s estimated degree, the discrepancy indicates the presence of connections to unobserved (hidden) nodes. By quantifying these discrepancies across the network, we further infer the number and connectivity of hidden components.

The core insight of our approach is that the spectral strength of a node reflects the cumulative effect of its interactions. By interpreting perturbation-induced spectral variations as footprints of connectivity, we transform local spectral observations into topological information. This strategy naturally integrates network reconstruction and hidden node identification into a single coherent pipeline—removing the need for model-specific assumptions or full-state observability.

\subsubsection{Degree Estimation from Spectral Strength}

To infer the connectivity structure of a network from limited observations, we begin by estimating the integer-valued degree of each observable node based solely on its dynamical behavior. This subsection presents a frequency-domain approach that translates node-level payoff time series into degree estimates, leveraging the empirically validated linear relationship between spectral strength and node degree (\( S_i \propto k_i \)).

The input to this degree estimation algorithm is the time series of observable node payoffs, with the output being their corresponding integer-valued degree distribution. The core idea is that each node’s spectral strength can be expressed as the product of its degree and a uniform scaling factor across nodes. We therefore iteratively search for the optimal scaling factor that minimizes the total squared error between the predicted and observed strength values.

\begin{algorithm}[htbp]
\caption{Estimate Integer Degree Sequence from Spectral Strength}
\label{alg:optimal_degree_from_strength}
\begin{algorithmic}[1]
\Require Observable node payoff time series \(\mathcal{T}\)
\Ensure Estimated integer degree sequence \(\mathbf{k}\)
\State Compute spectral strengths \(S\) from \(\mathcal{T}\) using DFT
\State Extract non-zero strengths to form \(S_{nz}\)
\If{\(S_{nz}\) is empty}
    \State \Return zero vector of length \(\lvert S\rvert\)
\EndIf
\State Determine median strength \(S_{med}\) from \(S_{nz}\)
\State Initialize \(\mathbf{k}\gets \mathbf{0}\), \(\epsilon_{min}\gets \infty\)
\For{\(k_c = 1\) to \(\lvert S\rvert - 1\)}
    \State Compute scaling factor \(S_p \gets S_{med}/k_c\)
    \ForAll{node \(i\)}
        \State Compute tentative degree \(\tilde{k}_i \gets \mathrm{round}(S_i/S_p)\)
        \State Estimate node strength \(\hat{S}_i \gets \tilde{k}_i \cdot S_p\)
    \EndFor
    \State Compute total squared error \(\epsilon \gets \sum_i (S_i - \hat{S}_i)^2\)
    \If{\(\epsilon < \epsilon_{min}\)}
        \State Update \(\mathbf{k}\gets \{\tilde{k}_i\}\), \(\epsilon_{min}\gets \epsilon\)
    \Else
        \State \Return \(\mathbf{k}\)
    \EndIf
\EndFor
\State \Return \(\mathbf{k}\)
\end{algorithmic}
\end{algorithm}

The estimation procedure (Algorithm~\ref{alg:optimal_degree_from_strength}) proceeds as follows. We first discard all zero entries from the strength sequence and compute the median of the remaining values, denoted \( S_{med} \), which serves as a robust reference strength. For each candidate reference degree \( k_c \in \{1, 2, \dots, |S| - 1\} \), we compute a corresponding scaling factor \( S_p = S_{med}/k_c \). This candidate slope is applied across all nodes to infer tentative degrees via \( \hat{k}_i = \mathrm{round}(S_i / S_p) \), followed by the computation of predicted strengths \( \hat{S}_i = \hat{k}_i \cdot S_p \). The quality of each candidate is evaluated by the total squared error \( \sum_i (S_i - \hat{S}_i)^2 \), and the degree sequence yielding the minimum error is retained as the final estimate.

This algorithmic formulation offers several advantages. It operates without requiring knowledge of the underlying network topology or nodal strategies, relying solely on observable dynamical signals. Its computational complexity scales linearly with the number of candidate degrees and is inherently parallelizable across nodes. The use of spectral features ensures robustness against noise, short time series, and local fluctuations. In addition, the fitting error exhibits a unimodal trend as \( k_c \) increases—first decreasing and then increasing—enabling early termination of the search to improve efficiency.

\subsubsection{Network Topology Inference and Hidden Node Detection}

This subsection presents a unified framework for inferring network topology and estimating hidden node connections using local perturbations. The core idea is to reconstruct each node’s neighborhood by identifying changes in spectral strength caused by suppressing its dynamic behavior. This mechanism leverages the locality of interactions in evolutionary games—each node interacts only with its direct neighbors. When a node is detached from the game dynamics, only its neighbors experience structural change, reflected as a noticeable drop in their spectral strength. These perturbation-induced strength reductions provide a direct signal for neighbor identification.

The overall inference procedure is as follows. Given the payoff time series of all observable nodes, we first estimate the reference degree sequence under unperturbed dynamics using Algorithm~\ref{alg:optimal_degree_from_strength}. We then iteratively suppress the dynamics of nodes whose neighborhoods have not yet been fully determined. For each target node, we recalculate the strength sequence from the perturbed time series and identify its neighbors based on strength drops, as detailed in Algorithm~\ref{alg:reconstruction_with_hidden}. If the number of inferred neighbors matches the reference degree, the node is assumed not to be connected to any hidden nodes. If fewer neighbors are identified than expected, the difference is attributed to hidden links. To reduce unnecessary computations, we skip perturbing nodes that have already accumulated the number of neighbors equal to their reference degree, as their full neighborhood has already been revealed through previous interventions.

Further analysis of hidden connectivity is achieved by maintaining a hidden link count vector \( L \). If \( L \) is a zero vector, no hidden nodes are present. If all nonzero entries in \( L \) equal one, the network likely contains a single hidden node—this matches prior assumptions in existing literature \citep{deng_yang_huang_wu_2022} \citep{hiddenGeoSpatial} \citep{shen_wang_fan_di_lai_2014}. If any \( L(i) > 1 \), multiple hidden nodes are required. In simple graphs without multi-edges, the lower bound on the number of hidden nodes equals the maximum entry in \( L \), while the upper bound equals the sum of all hidden link counts, assuming each connection links to a distinct hidden node. In other words, the lower bound corresponds to a scenario where each hidden node connects to all observable nodes with external links, while the upper bound assumes that each external edge originates from a distinct hidden node.

It is important to note that this framework critically depends on the linearity between spectral strength and node degree. Without this assumption, it would be difficult to distinguish perturbation-induced effects from endogenous noise or strategy fluctuations. 

\begin{algorithm}[htbp]
\caption{Network Structure Reconstruction with Hidden Link Estimation}
\label{alg:reconstruction_with_hidden}
\begin{algorithmic}[1]
\Require Time series \(\mathcal{T}\); number of nodes \(N\); dynamics \(\mathcal{PDG}\)
\Ensure Adjacency matrix \(M\); hidden links count vector \(L\)
\State Initialize \(M \gets \emptyset\), \(L \gets \mathbf{0}\in\mathbb{R}^N\)
\State Compute reference degree sequence \(k^{ref}\) from \(\mathcal{T}\) using Algorithm~\ref{alg:optimal_degree_from_strength}
\State Initialize known neighbors \(\mathcal{N}_i \gets \emptyset\) for each node \(i\)
\For{\(i = 1\) to \(N\)}
    \If{\(\lvert \mathcal{N}_i\rvert = k^{ref}(i)\)}
        \State \textbf{continue}
    \EndIf
    \State Suppress node \(i\)'s dynamics in \(\mathcal{PDG}\) to generate new time series \(\mathcal{T}_i\)
    \State Recompute degree sequence \(k^{new}\) from \(\mathcal{T}_i\)
    \State Identify neighbors \(\mathcal{N}_i^{new} \gets \{\,j \mid k^{new}(j) < k^{ref}(j)\}\)
    \State Update \(M(i,j)=1\) and \(M(j,i)=1\) for all \(j\in\mathcal{N}_i^{new}\)
    \If{\(\lvert \mathcal{N}_i^{new}\rvert < k^{ref}(i)\)}
        \State Set \(L(i)\gets k^{ref}(i) - \lvert \mathcal{N}_i^{new}\rvert\)
    \EndIf
\EndFor
\State \Return \(M, L\)
\end{algorithmic}
\end{algorithm}

\subsection{Evaluation Metrics}

To systematically assess the performance of network reconstruction, we adopt both machine learning-inspired classification metrics and network-specific accuracy measures. These complementary indicators facilitate cross-disciplinary interpretability while preserving the rigor of topological evaluation.

We begin with two widely used metrics in binary classification tasks: the \textit{true positive rate} (TPR) and \textit{false positive rate} (FPR). These are defined as:
\begin{equation}
\mathrm{TPR} = \frac{\mathrm{TP}}{\mathrm{TP} + \mathrm{FN}}, 
\quad
\mathrm{FPR} = \frac{\mathrm{FP}}{\mathrm{FP} + \mathrm{TN}},
\end{equation}
where TP, FP, TN, and FN denote true positives, false positives, true negatives, and false negatives, respectively. TPR quantifies the proportion of actual links correctly identified, while FPR captures the proportion of nonexistent links erroneously inferred.

To align with conventions in network science, we also report the \textit{successful reconstruction rate of existing links} (SREL) and \textit{non-existing links} (SRNL), defined as:
\begin{align}
\mathrm{SREL} &= \frac{\text{correctly reconstructed existing links}}
{\text{actual existing links}}\,, \nonumber\\
\mathrm{SRNL} &= \frac{\text{correctly identified non-existing links}}
{\text{actual non-existing links}}.
\end{align}

These metrics directly reflect reconstruction fidelity at the structural level. Mathematically, we note that \( \mathrm{SREL} = \mathrm{TPR} \) and \( \mathrm{SRNL} = 1 - \mathrm{FPR} \), yet they offer intuitive clarity when communicating topological accuracy.

In practice, we report average SREL and SRNL values across multiple independent realizations for each experimental setting. This averaging mitigates the impact of stochastic fluctuations in the dynamics and yields more robust estimates of reconstruction performance. Given the inherent sparsity of real-world networks—where absent links far outnumber existing ones—treating the two link classes separately ensures an unbiased and representative evaluation of each method’s performance.

This evaluation framework is consistent with prior studies in the field, such as Ref.~\citep{physrevx.1.021021}, and enables transparent benchmarking against alternative inference techniques.

\subsection{Benchmark Methods}

To evaluate the effectiveness of the proposed DFT-based method for network reconstruction, we compare its performance against several widely used baseline approaches. These methods span classical statistical inference techniques and information-theoretic tools, providing diverse perspectives on structure recovery from time series data.

Several recent advanced methods are excluded due to fundamental incompatibilities with our experimental design. For instance, the ARNI model~\citep{casadiego2017model} requires known functional forms and continuous dynamics, which are absent in our discrete-time evolutionary games. Similarly, the Neural Relational Inference (NRI) model~\citep{kipf2018neural} exhibits scalability limitations and was found to be unstable for networks exceeding 50 nodes.

Furthermore, compressed sensing-based methods assume access to individual strategy trajectories, whereas our framework only observes node-level payoffs. As such, these methods are excluded from direct comparison in our evaluations.

We adopt the following four representative baseline methods as interpretable and practical alternatives:

\begin{itemize}
    \item \textit{Granger Causality (GC)} \citep{GrangerC1969}: Quantifies the causal influence of one time series on another by comparing prediction errors from autoregressive models, with and without the inclusion of the candidate input. Each node pair is scored via the log-ratio of error variances.

    \item \textit{Correlation Matrix (CM)} \citep{CorrelationMatrix2010}: Measures linear statistical dependence between node time series. Despite its simplicity, CM remains widely used due to its ease of implementation and interpretability.

    \item \textit{Mutual Information (MI)} \citep{butteMutualInformation2000}: Captures nonlinear dependencies between pairs of time series by measuring reductions in uncertainty. It provides a model-free estimate of pairwise association strength rooted in information theory.

    \item \textit{Maximum Likelihood Estimation (MLE)}: Estimates interaction strengths by maximizing the likelihood of observed node trajectories under parametric models, enabling inference of weighted connectivity.
\end{itemize}

All baseline methods are implemented using the \texttt{netrd} Python library~\citep{mccabe2020netrd}, which provides consistent interfaces and default settings to ensure fair and reproducible evaluation. By benchmarking against this diverse set of methods, we demonstrate the robustness and advantages of our spectral inference framework across varied topologies and dynamical regimes.

\subsection{Experimental Setup}

To validate the effectiveness of our proposed spectral inference framework, we conduct extensive numerical experiments using synthetic evolutionary game dynamics on various network topologies. All experiments were performed on a workstation equipped with an Intel Core i7-13700F CPU (2.10\,GHz) and 16\,GB RAM.

The experimental protocol is as follows:

\begin{itemize}
    \item Each node in the network acts as a game agent, updating strategies and accumulating payoffs as defined in Section~\ref{sec:behavior_series}. The payoff matrix parameters are set to \( r = 3 \), \( s = 0 \), \( t = 5 \), and \( p = 1 \) unless otherwise stated.
    \item All nodes initialize with zero payoff. Strategy updates follow the Fermi rule, with a high temperature parameter (\(10^8\)) to simulate near-random behavior. This configuration ensures that the payoff dynamics primarily reflect structural interactions.
    \item To mitigate stochastic variability, all results are averaged over multiple independent realizations, ensuring statistical robustness and generalizability.
\end{itemize}

\section{Data}
\label{sec:data}
This section presents the network datasets and dynamic processes used to evaluate our network reconstruction framework. We include both real-world and synthetic networks to ensure the generalizability of our approach across different structural properties. In addition, we provide a detailed description of the discrete-time evolutionary game dynamics used to generate node-level payoff sequences, as well as the stochastic strategy update mechanism. Together, these form the foundation for applying the proposed frequency-domain inference framework.

\subsection{Empirical and Synthetic Network Datasets}
To comprehensively evaluate the proposed method, we apply it to both empirical networks drawn from real-world systems and synthetic networks with well-controlled topologies. 

The empirical networks span social, biological, and technological systems. These include:
\begin{itemize}
\item \textbf{Animal interaction} \citep{animalsocial}: Sheep: Social status was determined by assembling a win-loss matrix based on the outcome of agonistic interactions. The winner of a dominance fight, involving a series of butts and clashes, was recorded as winning one interaction. Dolphin \citep{Lusseau_Newman_2004}:This is a social network of bottlenose dolphins. The nodes are the bottlenose dolphins (genus Tursiops) of a bottlenose dolphin community living off Doubtful Sound, a fjord in New Zealand (spelled fiord in New Zealand). An edge indicates a frequent association. The dolphins were observed between 1994 and 2001. Songbird: A machine learning algorithm was applied to identify clusters of detections on feeders. Next, the network was generated based on patterns of co-occurrence by individuals in the same feeding events. Associations between birds were defined using the simple ratio index.
\item \textbf{Online social platforms} \citep{twitter}: Twitter: Nodes are twitter users and edges are retweets. These were collected from various social and political hashtags. Retweet is a retweet and mentions network from the UN conference held in Copenhagen. The data was collected over a two week period.
\item \textbf{Email communication} \citep{nr}: Enron is a network based on email communications.
\item \textbf{Brain connectivity} \citep{nr}: Mouse visual cortex: Edges represent fiber tracts that connect one vertex to another.
\item \textbf{Infrastructure} \citep{infUSAir}: USAir transportation: A network based on airport connections.
\item \textbf{Crime} \citep{nr}: Crime: This network contains persons who appeared in at least one crime case as either a suspect, a victim, a witness or both a suspect and victim at the same time. A left node represents a person and a right node represents a crime. An edge between two nodes shows that the left node was involved in the crime represented by the right node. Madrid Train Bombing Network \citep{memon_larsen_hicks_harkiolakis_2008}: This network contains contacts between suspected terrorists involved in the train bombing of Madrid on March 11, 2004 as reconstructed from newspapers. A node represents a terrorist and an edge between two terrorists shows that there was a contact between the two terrorists. This includes friendship and co-participating in training camps or previous attacks.

\end{itemize}
These networks offer broad variation in size, degree distribution, and assortativity, providing a robust testbed for inference tasks.

\begin{table}
    \caption{Topological statistics of empirical networks. These statistics include the number of nodes (\(N\)), the number of links (\(E\)), average degree (\(\langle k\rangle\)), max degree (\(k_{\max}\)), min degree (\(k_{\min}\)), and degree assortativity (\(DA\)) \citep{newman_2003,foster_foster_grassberger_paczuski_2010}.}
    \centering
    \begin{tabular}{llllllll}
    \toprule
    Network               & Short form        & \(N\) & \(E\)   & \(\langle k\rangle\) & \(k_{\max}\) & \(k_{\min}\) & \(DA\)     \\
    \midrule
    Sheep Dominance       &    Sheep          &  28   &  235    & 16.79                &  23          &  4           & -0.0040 \\
    Dolphin Social        &    Dolphin        &  62   &  159    &  5.13                &  12          &  1           & -0.0436 \\
    Madrid Train Bombing  &    Madrid         &  64   &  243    &  7.59                &  29          &  1           &  0.0295 \\
    ca-sandi auths        &    Sandi          &  86   &  124    &  2.88                &  12          &  1           & -0.2558 \\
    rt-retweet            &    Retweet        &  96   &  117    &  2.44                &  17          &  1           & -0.1792 \\
    aves-songbird-social  &    Songbird       & 110   & 1027    & 18.67                &  56          &  1           &  0.0003 \\
    email-enron-only      &    Email          & 143   &  623    &  8.71                &  42          &  1           & -0.0195 \\
    mouse visual cortex 2 &    Mouse          & 193   &  214    &  2.22                &  31          &  1           & -0.8447 \\
    inf USAir 97          &    USAir          & 332   & 2126    & 12.80                & 139          & 1            & -0.2079 \\
    rt-twitter-copen      &    Twitter        & 761   & 1031    & 2.71                 & 37           & 1            & -0.0993 \\
    ia-crime-moreno       &    Crime          & 829   & 1476    & 3.56                 & 25           & 1            & -0.1635 \\
    \bottomrule
    \end{tabular}
    \label{tab:table1}
\end{table}

Table~\ref{tab:table1} summarizes the topological statistics of the empirical networks, including the number of nodes and edges, average and extreme degrees, and degree assortativity.

For synthetic networks, we generate them using three canonical models— Erdős–Rényi (ER) \citep{erdos_rényi_1986}, Barab{\'a}si–Albert (BA) \citep{Barabasi_Albert_1999} and Watts–Strogatz (WS) \citep{watts1998collective}—with controlled network size, edge density, and average degree. Details of all synthetic networks used in our reconstruction experiments are provided in Table~\ref{tab:density_nets}.

\begin{table}[ht]
  \centering
  \caption{\textbf{Structural properties of synthetic networks used in reconstruction experiments.} All networks are designed to have approximately the same average degree. For networks of the same size, the edge density is also controlled to be comparable across ER, BA, and WS models.}  \label{tab:density_nets}
  \begin{tabular}{lrrr}
    \toprule
    Network & Size  & Density & Average degree  \\
    \midrule
    BA-100  & 100   & 0.1139  & 11.28           \\
    ER-100  & 100   & 0.1133  & 11.22           \\
    WS-100  & 100   & 0.1212  & 12.00           \\
    BA-200  & 200   & 0.0585  & 11.64           \\
    ER-200  & 200   & 0.0583  & 11.60           \\
    WS-200  & 200   & 0.0603  & 12.00           \\
    BA-400  & 400   & 0.0296  & 11.82           \\
    ER-400  & 400   & 0.0297  & 11.84           \\
    WS-400  & 400   & 0.0301  & 12.00           \\
    BA-600  & 600   & 0.0198  & 11.88           \\
    ER-600  & 600   & 0.0188  & 11.25           \\
    WS-600  & 600   & 0.0200  & 12.00           \\
    \bottomrule
  \end{tabular}
\end{table}

\subsection{Evolutionary Game Dynamics on Networks}\label{sec:behavior_series}

To simulate the behavior of agents on networks, we adopt the Prisoner's Dilemma game (PDG) \citep{nowak_may_1992} as the underlying interaction model. Each node acts as a game-playing agent that repeatedly interacts with its neighbors. At each round, a node selects a strategy—either cooperation (C) or defection (D)—and receives a payoff based on its own and its neighbors' strategies. The strategy vectors are encoded as \(\mathbf{S}(C)=(1,0)^\top\) and \(\mathbf{S}(D)=(0,1)^\top\), where the notation \(\top\) indicates the transpose. The general form of the PDG payoff matrix is defined as
\begin{equation}
\mathbf{P}_{\mathrm{PDG}} = \begin{pmatrix} r & s \\ t & p \end{pmatrix},
\end{equation}

where \( r \) is the payoff for mutual cooperation, \( s \) for cooperating against a defector, \( t \) for defecting against a cooperator, and \( p \) for mutual defection.

Each node plays the game with all its neighbors and accumulates the total payoff. At each round \( t \), the total payoff of node \( i \) is computed as
\begin{equation}
P_i(t) = \sum_{j \in \Gamma_i} \mathbf{S}_i^\top \mathbf{P} \mathbf{S}_j,
\end{equation}
where \( \Gamma_i \) is the set of neighbors of node \( i \), and \( \mathbf{S}_i \), \( \mathbf{S}_j \) denote the strategies of nodes \( i \) and \( j \), respectively. The sequence \( \{ P_i(t) \}_{t=1}^T \) forms the payoff time series used for structure inference.

\subsection{Strategy Update Mechanism}

After each game round, nodes update their strategies based on the payoffs of their neighbors. We adopt the Fermi rule \citep{szabó_tőke_2002}, a widely used stochastic update mechanism in evolutionary dynamics. Specifically, node $i$ randomly selects a neighbor $j$, and adopts $j$'s strategy with probability
\begin{equation}
W(\mathbf{S}_i \leftarrow \mathbf{S}_j) = \frac{1}{1 + \exp \left[ (P_i(t) - P_j(t))/\kappa \right]},
\end{equation}
where $P_i(t)$ and $P_j(t)$ are the payoffs of nodes $i$ and $j$ at round $t$, and $\kappa$ is a positive parameter that controls the level of stochasticity in the update process. When $\kappa \to 0$, the update becomes nearly deterministic: node $i$ is almost certain to adopt the strategy of node $j$ if $P_j(t) > P_i(t)$. Conversely, as $\kappa \to \infty$, the update becomes increasingly random, approaching equal probability regardless of payoffs.

In our experiments, we investigate the behavior of the proposed method under both random and rational strategy updating regimes. We vary the parameter $\kappa$ to simulate these different dynamics: a large value (e.g., $\kappa = 10^8$) approximates random updates, while a small value (e.g., $\kappa = 10^{-8}$) leads to highly rational imitation dynamics. This flexibility allows us to assess the robustness of the payoff–structure relationship across varying levels of behavioral noise. The resulting payoff time series, shaped by these dynamics, form the basis for inferring network structure using our DFT-based framework.

\section{Discussion}
\label{sec:discussion}

In this study, we identified and validated a robust linear relationship between the spectral strength of nodes' payoff dynamics and their structural degree (\( S \propto k \)), providing a theoretical foundation for inferring network topology from behavioral observations. This relationship emerges naturally due to the intrinsic dynamics of evolutionary games, where a node’s payoff is the cumulative outcome of interactions with its neighbors. Higher-degree nodes interact with more partners, accumulating perturbations reflected directly in the frequency domain. The linearity is further reinforced by the linear operation of the DFT, enabling a straightforward proportionality between dynamic fluctuations and structural connectivity.

The consistency of this linear scaling across diverse network types (including model networks such as BA, ER, and WS, as well as empirical networks) and varying dynamical settings (high and low selection intensities) underscores its universality, suggesting that similar principles may apply broadly to other nonlinear dynamical systems. In particular, under stochastic Fermi updates, the bifurcation into cooperation and defection states clearly illustrates how strategy-specific payoffs modulate the linear coefficient without altering the underlying structural proportionality. Conversely, under highly rational dynamics (low selection intensity), all nodes converge to a single Nash equilibrium, removing payoff-induced heterogeneity but preserving structural scaling. This interpretation elucidates why the spectral strength reliably encodes structural information, independent of transient behavioral variations.

Compared to conventional methods—such as Granger causality, mutual information, correlation matrices, and maximum likelihood estimation—our frequency-domain framework consistently achieves superior reconstruction accuracy. Its strengths include robustness to noise, minimal model assumptions, and the ability to simultaneously identify hidden nodes and estimate their external connectivity. The method's high accuracy across synthetic and empirical networks indicates its practical applicability, particularly for analyzing partially observable systems.

Nonetheless, the proposed framework has inherent limitations. It specifically leverages evolutionary game dynamics, and its applicability to alternative dynamical processes remains untested. The necessity of dynamic interventions to isolate local neighborhood structures may also limit usability in certain real-world contexts. Additionally, reconstruction accuracy remains sensitive to the length and quality of available time series data, which poses practical constraints for datasets that are short or irregularly sampled.

\section{Conclusion}
\label{sec:conclusion}

This study uncovered a fundamental linear scaling law between the spectral strength of node-level payoff dynamics and network topology in evolutionary games, establishing a bridge between local dynamic observables and global structural features. Utilizing this insight, we developed a robust inference framework based on the DFT, enabling precise network topology reconstruction and reliable detection of hidden nodes and external connections, without requiring prior knowledge of the network or node-level strategies.

Extensive evaluations demonstrated that our method consistently outperformed traditional techniques across various synthetic and empirical networks, maintaining accuracy, scalability, and robustness even as network size increased. Moreover, our approach effectively identified and quantified hidden structures, providing valuable insights into complex systems characterized by partial observability.

These findings significantly advance our understanding of how structural information manifests in nonlinear dynamics, establishing a foundation for accurate network inference from local behavioral signals. This opens avenues for a wide range of real-world applications, including the identification of latent social ties—such as covert links in criminal or terrorist organizations, the reconstruction of gene regulatory networks from gene expression data, the mapping of neural connectomes based on neuronal activity time series, and fault localization in power grids using voltage or frequency fluctuations. Future work should explore the generality of the observed spectral linearity beyond evolutionary games, develop inference methods that remain effective under limited or noisy observations, and further enhance capabilities for detecting and characterizing hidden components in partially observable systems.

\backmatter




\bmhead{Acknowledgements}

This work was supported by the National Natural Science Foundation of China (Grant Nos. T2293771, 62503447), the STI 2030 Major Projects (Grant No. 2022ZD0211400), the China Postdoctoral Science Foundation (Grant No. 2024M763131), the Postdoctoral Fellowship Program of CPSF (Grant No. GZC20241653), and the New Cornerstone Science Foundation through the XPLORER PRIZE.





\bmhead{Funding}

Not applicable

\bmhead{Competing interests}

The authors declare no competing interests.

\bmhead{Ethics approval and consent to participate}

Not applicable

\bmhead{Consent for publication}

Not applicable

\bmhead{Data availability}

The data and code supporting the findings of this study are publicly available at: \textcolor{red}{\url{https://github.com/XXX/XXX}}.

\bmhead{Materials availability}

Not applicable

\bmhead{Code availability}

The data and code supporting the findings of this study are publicly available at: \textcolor{red}{\url{https://github.com/XXX/XXX}}.

\bmhead{Author contribution}

X.L., T.F. and L.L. conceived and designed the research. X.L. developed the initial conceptual framework, performed numerical experiments, and drafted the manuscript. T.F. refined the theoretical framework, optimized experimental designs, and substantially revised and expanded the manuscript through comprehensive restructuring and scholarly enhancement. T.F. and L.L. supervised the research and critically reviewed and edited the manuscript. All authors reviewed and approved the final manuscript.





\bibliography{references}

\clearpage
\begin{appendices}

\section*{Supplementary Information}\label{Sec_SI}


\begin{figure}[htbp]
	\centering
        \includegraphics[scale=0.6]{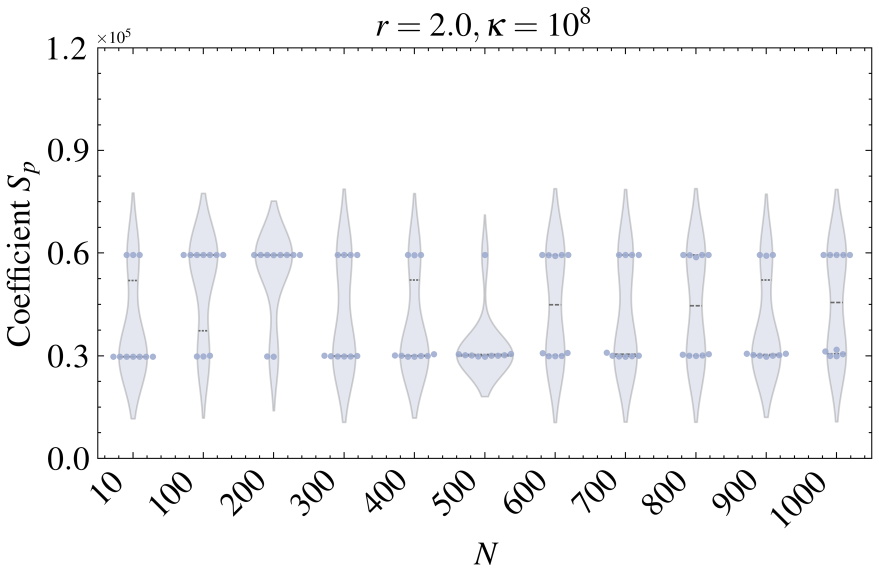}
	\caption{\textbf{Distribution of linear coefficients $S_p$ under strong selection intensity ($\kappa = 10^8$) with $r = 2$.} Each violin reflects 10 independent simulation trials. The fitted intercepts exhibit a bimodal distribution aligned with the payoff ratio ($r:p = 2\!:\!1$). All results are obtained with time series length $T = 10{,}000$.}
	\label{fig:Supple_coeff_N_b}
\end{figure}

\begin{figure}[htbp]
	\centering
        \includegraphics[scale=0.6]{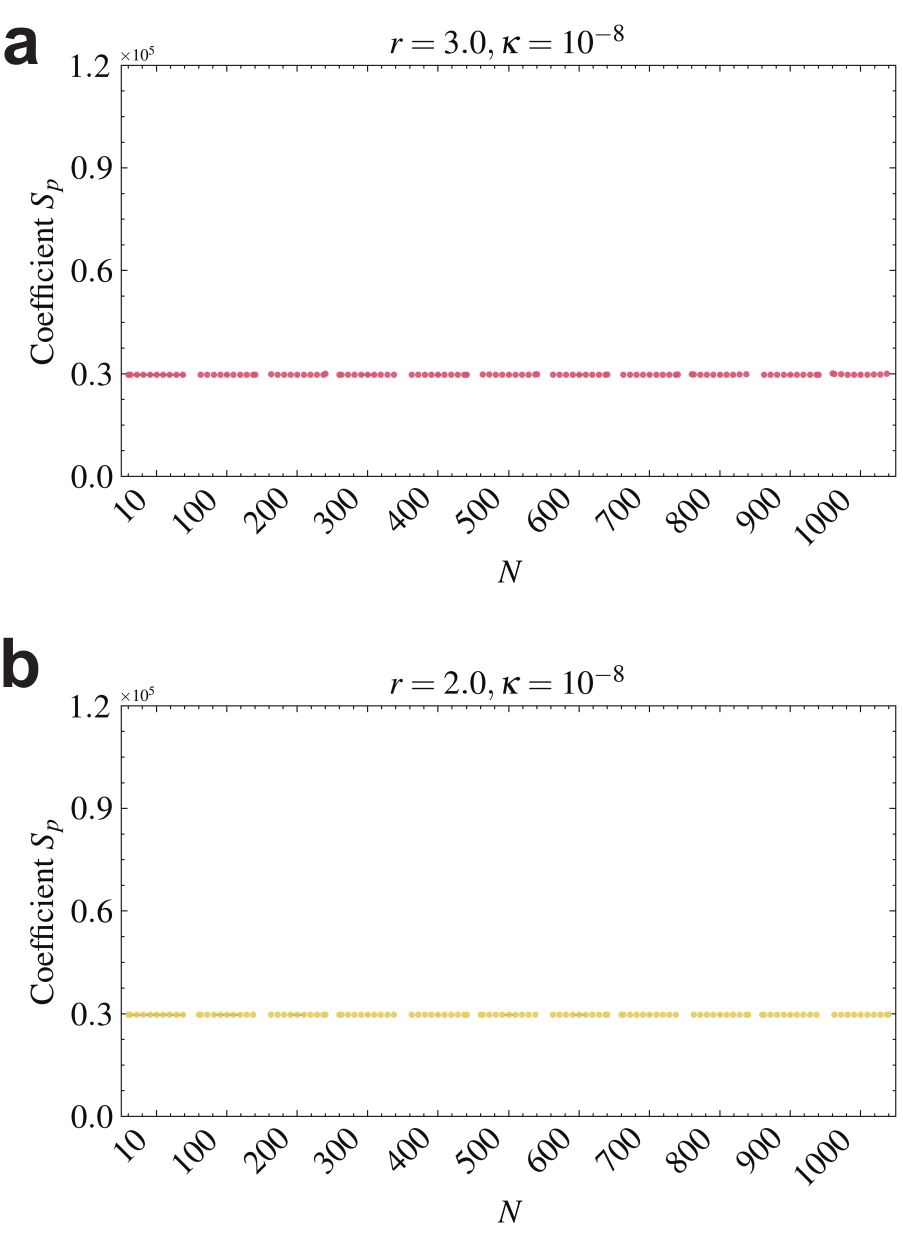}
	\caption{\textbf{Comparison of fitted linear coefficients $S_p$ under weak selection intensity ($\kappa = 10^{-8}$) and varying cooperation payoffs.}  
    Distributions of the fitted linear coefficients $S_p$ for (a) \( r = 3.0 \) and (b) \( r = 2.0 \). Each point corresponds to the fitted $S_p$ from one of 10 independent simulation trials at a given network size $N$. Across all sizes and both payoff settings, the $S_p$ remain nearly identical, confirming the robustness of the strength–degree linearity to changes in cooperation incentives. All results are obtained with $p = 1$ and time series length $T = 10{,}000$.}
	\label{fig:Supple_compare_r}
\end{figure}

\begin{figure}[htbp]
	\centering
	\includegraphics[scale=0.5]{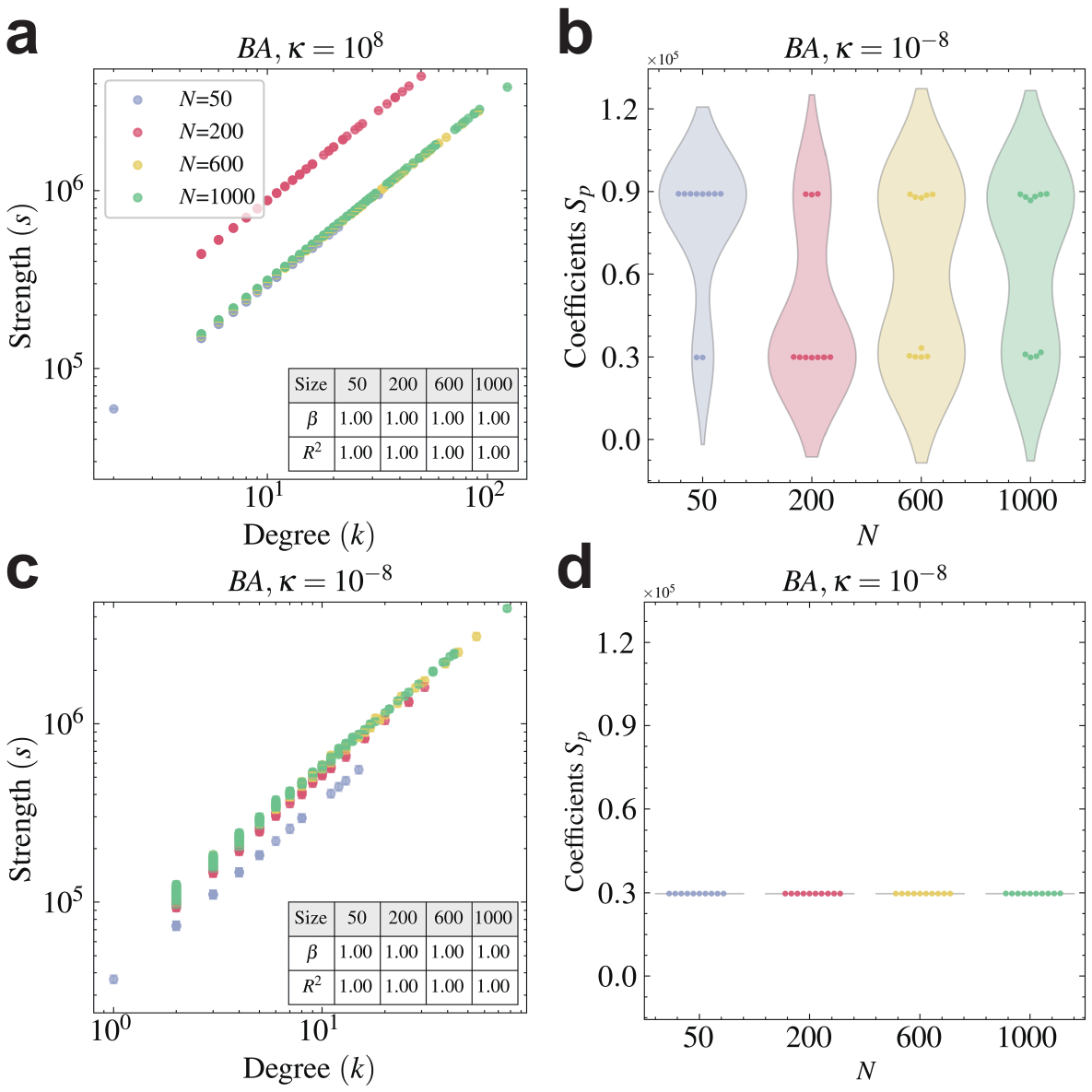}
	\caption{\textbf{Empirical validation of strength--degree linearity on BA networks of varying sizes.} 
        (a) and (c) display the relationship between node strength and degree on BA networks of different sizes ($N = 50$, $200$, $600$, $1000$), under strong ($\kappa = 10^8$) and weak ($\kappa = 10^{-8}$) selection intensities, respectively. Each point represents a node; colors distinguish network sizes. The fitted model $S \propto k^\beta$ is applied in log--log space, with fitted slopes $\beta$ and $R^2$ scores summarized in the insets. 
        (b) and (d) show the distributions of fitted linear coefficients $S_p$ (intercepts from panels (a) and (c), respectively), aggregated over 10 independent trials for each network size.
        Under strong selection, neutral drift yields stochastic convergence to either cooperation or defection in each trial, producing two parallel lines in (a) and a bimodal distribution in (b), aligned with the payoff ratio ($r:p = 3\!:\!1$). 
        Under weak selection, strategy convergence becomes deterministic (toward defection), collapsing the strength--degree relation to a single trend and eliminating bimodality, as shown in (c) and (d).
        All simulations use $r = 3$, $p = 1$, and time series length $T = 10{,}000$.}
	\label{fig:linear_BA}
\end{figure}

\begin{figure}[htbp]
	\centering
	\includegraphics[scale=0.5]{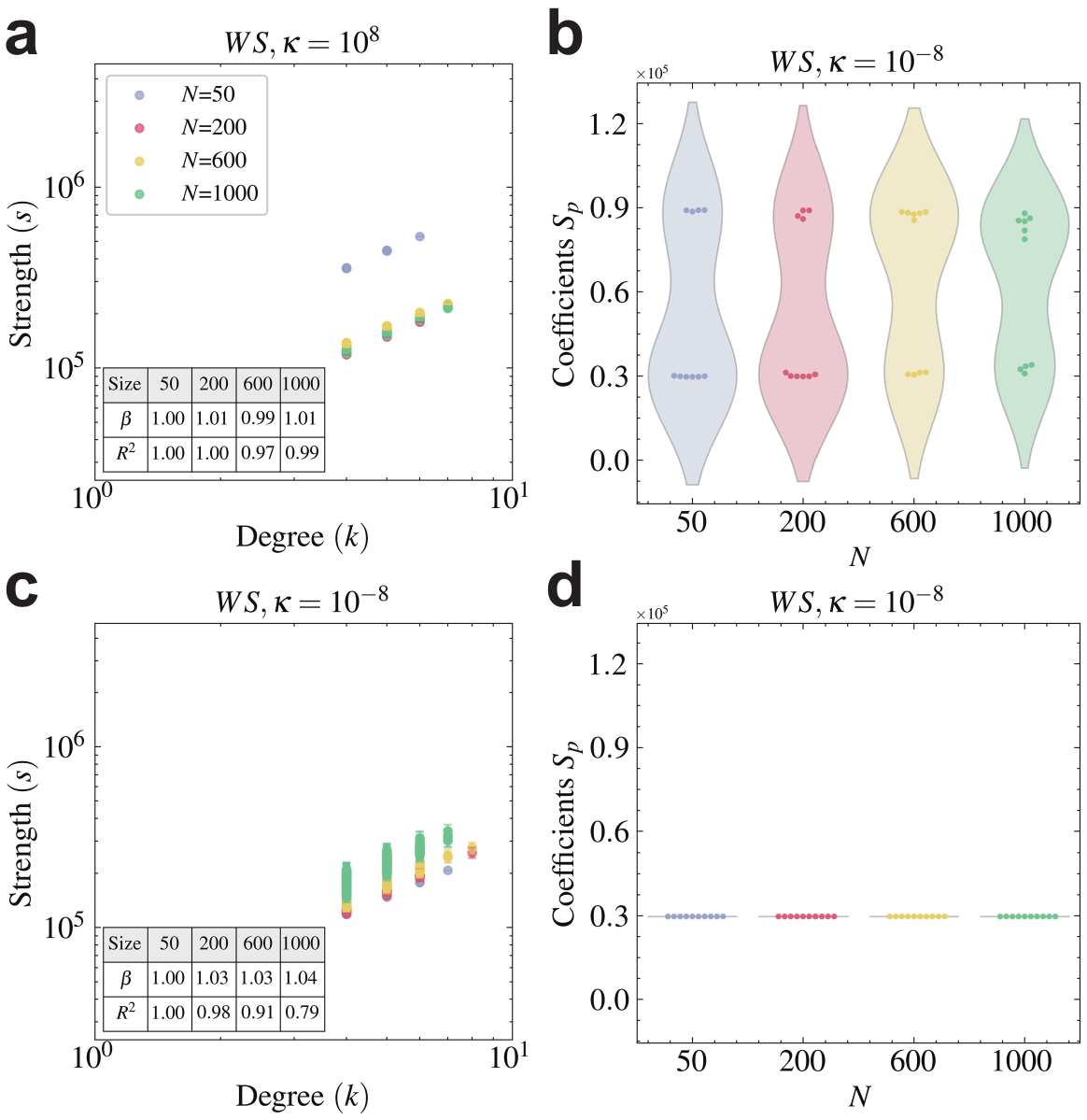}
	\caption{\textbf{Empirical validation of strength--degree linearity on WS networks of varying sizes.} 
        (a) and (c) display the relationship between node strength and degree on WS networks of different sizes ($N = 50$, $200$, $600$, $1000$), under strong ($\kappa = 10^8$) and weak ($\kappa = 10^{-8}$) selection intensities, respectively. Each point represents a node; colors distinguish network sizes. The fitted model $S \propto k^\beta$ is applied in log--log space, with fitted slopes $\beta$ and $R^2$ scores summarized in the insets. 
        (b) and (d) show the distributions of fitted linear coefficients $S_p$ (intercepts from panels (a) and (c), respectively), aggregated over 10 independent trials for each network size.
        Under strong selection, neutral drift yields stochastic convergence to either cooperation or defection in each trial, producing two parallel lines in (a) and a bimodal distribution in (b), aligned with the payoff ratio ($r:p = 3\!:\!1$). 
        Under weak selection, strategy convergence becomes deterministic (toward defection), collapsing the strength--degree relation to a single trend and eliminating bimodality, as shown in (c) and (d).
        All simulations use $r = 3$, $p = 1$, and time series length $T = 10{,}000$.}
	\label{fig:linear_WS}
\end{figure}




\end{appendices}



\end{document}